# Angles are inherently neither length ratios nor dimensionless

Paul Quincey[a], Peter J Mohr[b] and William D Phillips[b]

[a] National Physical Laboratory, Hampton Road, Teddington, TW11 0LW, United Kingdom
[b] National Institute of Standards and Technology, Gaithersburg, MD 20899, USA

**Abstract**

The status of angles within The International System of Units (SI) has long been a source of controversy and confusion. We address one specific but crucial issue, putting the case that the idea of angles necessarily being length ratios, and hence dimensionless, is not valid. By making a clear distinction between the definition of a quantity and the process of making a measurement of that quantity, we show that the usual arguments for angles being length ratios are problematic. An appreciation of this point should clear away a major obstacle preventing a more proper treatment of the angle problem.

**Introduction**

One way to point out the odd status of angles within the SI is as follows: angle is measurable with a scale, on a protractor, in analogy to the way length is measureable with a ruler. It can therefore have units that are chosen by the user. This simple point is reinforced by the fact that multiple units for angle are used in practice. Angle measurements often use the degree, while the radian, which is used in mathematics and theoretical physics, is the SI unit. Other units, such as the grad, are used in specialised areas.

And yet angle is treated as being dimensionless in the SI, a status usually applied only to quantities that are simply counted, like the number of molecules in a container, or quantities that are clearly a ratio of two values of the same quantity, such as refractive index. The CGPM resolution that led to the SI position is, however, curiously worded: "to *interpret* … the radian and steradian as dimensionless derived units …" (our italics) (Resolution 8 of the 20[th] CGPM, 1995). They are therefore to be *treated* as dimensionless, which is not the same as their being *inherently* dimensionless. Previously, since its introduction in 1960, the SI had treated radians and steradians as supplementary units, a specially-created category separate from either base units or derived units. This highlights the difficulties of incorporating angles into the SI since it was first formalised.

Some earlier contributions to the issue can be found in Brinsmade 1936 and Brownstein 1997. Resilient divergent views remain, in roughly three camps, represented by: the SI Brochure (BIPM 2006 and 2018); Mohr and Phillips (2014), who suggest that angles should be treated as dimensional and that significant changes need to be made to the SI; and Quincey (2016) / Quincey and Brown (2017), who suggest that angles should continue to be treated as dimensionless within the SI, and that the associated problems can be avoided by recognizing that their status is a convention rather than a necessity. In fact, the diversity of viewpoints expressed in these and other works underscores the fact there is no single "correct" way to treat angles, but rather the goal should to be to provide a way that is most useful for the users of units.

It could be said explicitly that the status of angles as physical quantities, and of radians as units, is not decided by the SI in isolation. The status of any physical quantity is linked to far-reaching questions about the number of independent physical quantities, and the fundamental relationships between quantities. All coherent unit systems must incorporate this basic physics in one way or another. The form of the equations commonly used to express relationships between quantities can provide an additional constraint, which unit systems also need to take into account. The issue being addressed here is related to the fact that, in equations, the constant $\theta_N$ (as defined below) is generally taken to be equal to unity. The SI can make various choices about how it treats angles within this overall context.

When carefully examining units and their properties, it has become conventional to use the notation in which a quantity is explicitly written as a product of an ordinary number times a unit. In particular, a quantity may be expressed as $q = \{q\}[q]$, where $\{q\}$ is an ordinary number and $[q]$ is the unit to which that number corresponds. For example, a length $x$ of two metres would be expressed as $x = 2$ m, where $\{x\} = 2$ and $[x] = $ m. When quantities are combined, the numbers and units are combined separately following the conventional rules of algebra. For a velocity $v$, the algebraic rules give, for example, $v = 6$ m/2 s $= 3$ m/s. Angles within the SI often do not follow this convention, as explained in the next section. For example, $\theta$ is conventionally used to represent $\{\theta\}$, with $[\theta]$ implicitly understood to equal rad, when it appears as the argument of a trigonometric function.

It is not the aim of this paper to summarise all the points concerning angles made by the various camps, but to address what we see as a crucial issue concerning the treatment of angles as dimensionless, as implied by the text of the SI Brochure. We will do this by directly addressing several misconceptions that feature in discussions of the topic.

**Several misconceptions that arise in the discussion of angles**

Misconception: mathematics tell us that angle is a length ratio since $s = r\theta$

The first misconception, in effect, makes deductions about the status and unit of angle using as the starting point the familiar equation $s = r\theta$, whose terms are defined below. We argue against this because there are unstated assumptions built into the line of reasoning, as shown by the following considerations. Two equations for the arc length $s$ of a segment of a circle of radius $r$ are

$$s = \pi r\theta/180, \quad \text{where } \theta \text{ is in degrees,}$$

and
$$s = r\theta, \quad \text{where } \theta \text{ is in radians.}$$

Which of these is correct? Of course they are both correct, but they are both special cases, where the unit for angle is prescribed. Using the metrological notation above, they would be presented as

$$s = \pi r\{\theta\}/180, \quad \text{with } [\theta] = °,$$

and
$$s = r\{\theta\}, \quad \text{with } [\theta] = \text{rad}.$$

A general unit-independent equation[1] for arc length can be written as

$$s = r\theta/\theta_N.$$

---

[1] Bridgman (1922) page 37 - used the term "complete equations" for ones that remain unchanged for any choice of fundamental units. He pointed out that complete equations must always be used when inferring the dimensions of quantities, or the relationships between the units for different quantities, from an equation.

In this case, $\theta$ and $\theta_N$ can be in any unit for angle, providing it is the same for both, and $\theta_N$ is a constant[2], the angle subtended at a circle's centre by an arc of that circle when the length of the arc is equal to the radius, *i.e.,* 1 rad or $(180/\pi)°$. This equation can also be written as

$$\theta = \frac{s}{r}\theta_N$$

emphasising that this is a metrologically conventional equation where $\{\theta\} = s/r$ and $[\theta] = \theta_N$. Unlike the unit-dependent equations above, the general equations contain no information relating to whether or not angle is dimensionless.

It can now be seen that the mathematics does not tell us that "angle is *equal* to the ratio of arc length to radius", but rather that "angle is *proportional* to the ratio of arc length to radius." The proportionality constant $\theta_N$, which could be dimensional or simply a number, depends on our choice of units. As emphasised in the next section, it is important to note that the ratio provides a measurement of the angle (in radians), which should be distinguished from the quantity of the angle, which includes the factor $\theta_N$.

The equation $s = r\theta$ is simplified from the general equation by assuming both that $\theta$ is expressed in radians and that angle is dimensionless, so that the proportionality constant $\theta_N$ is set equal to the number one. Using this equation to "prove" that angle is dimensionless, and that the quantity of angle 1 rad is mathematically equivalent to the number one, is a circular argument.

Misconception: an angle is defined as the ratio of an arc length to a radius

The second misconception is closely related to the first, but does not rely on a specific equation. It is a more general statement that an angle is defined as the ratio of two lengths. The general equations in the previous section show that this statement is incorrect. Nevertheless, we find it useful to give an additional perspective. In particular, we show the statement is problematic by examining its logical extension to different units where traditional assumptions do not obscure the meaning.

Bathroom scales measure force – when we use them to measure our mass in kilograms we are taking the ratio of our weight (proportional to mass) to the weight of a kilogram. Does this show that mass is a force ratio, and therefore dimensionless? Of course not - it is a simply a normal measurement that compares an object with a standard.

We can measure an angle in radians as the ratio of an arc length on a circle (proportional to angle) to the arc length of a radian on the same circle, which happens to be the same as its radius. Does this show that angle is a length ratio? Again, no. It is just another measurement made by a comparison with a standard.

Mass is not *defined* as the ratio of two forces. A mass expressed in kilograms can be *measured* as the ratio of a weight to the weight of a kilogram. Similarly, an angle is not *defined* as the ratio of an arc length to a radius. An angle expressed in radians can be *measured* as the ratio of an arc length to a radius.

---

[2] This constant has the same dimension as angle, and should not be assumed *a priori* to be dimensionless. The N in the symbol $\theta_N$ refers to this being the natural unit of angle.

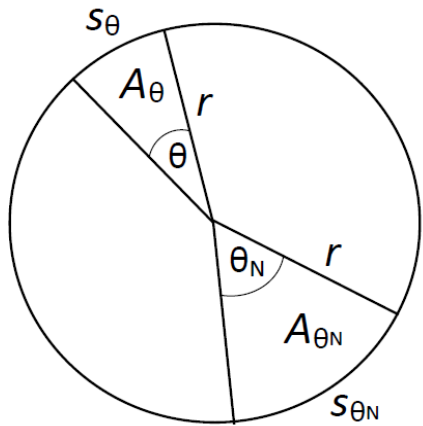

Figure 1   Some basic quantities relating to angles. The ratio of the two angles $\theta/\theta_N$ equals the ratio of their arc lengths $s_\theta/s_{\theta_N}$, and the ratio of the associated sector areas $A_\theta/A_{\theta_N}$. Because $\theta_N$ is 1 rad or $(180/\pi)°$, $s_{\theta_N} = r$ and $A_{\theta_N} = r^2/2$.

If length ratios were the only way to measure angles, there might be grounds for thinking that this represented some kind of special relationship. But it is not. For example, the areas $A$ of the sectors in Figure 1 are also proportional to angle, and so we can measure an angle as the ratio of its sector area to the sector area of our chosen angle unit in the same circle; this does not show that angle is an area ratio. Or we could measure angles by weighing the sectors shown, after cutting them from a uniform disc, for example made of paper; this does not show that angle is a mass ratio.

Most measurements of any quantity are a comparison between the unknown and a standard of the same quantity, like measuring length against a metre rule. If we never measured angle in this way, there might be grounds for treating angle differently. But of course this is exactly what we do when using a protractor, the most common method for measuring an angle, where we compare an unknown angle with the standard angles marked on the protractor.

Misconception: angle measurement requires length measurement

It might be argued that, whatever the preceding points imply about the nature of angles, it is not possible to measure an angle in practice without measurements of length, and this would justify the status of angle as a length ratio. However, a protractor can be made without length measurement rather easily.

Binary subdivisions of the angle represented by a straight line can be made using the well-known fact that angles can be repeatedly bisected geometrically with a compass and a straight edge. No length measurement is required. An even simpler method leading to a similar result is to use the following procedure:

We start with a piece of paper with a reference point (RP) marked near the centre.  We fold the paper so that the crease intersects the RP.  We fold the paper again with a new crease that includes the reference point and so that the previous crease on one side of the RP aligns with the previous crease on the other side of the RP.  This can be repeated (in principle) as many times as we wish. When the paper is unfolded, as in Figure 2, the creases are binary angular subdivisions of a complete rotation, which can readily be converted to degrees or radians by applying a simple conversion factor using the fact that one rotation is 360 degrees or $2\pi$ radians.

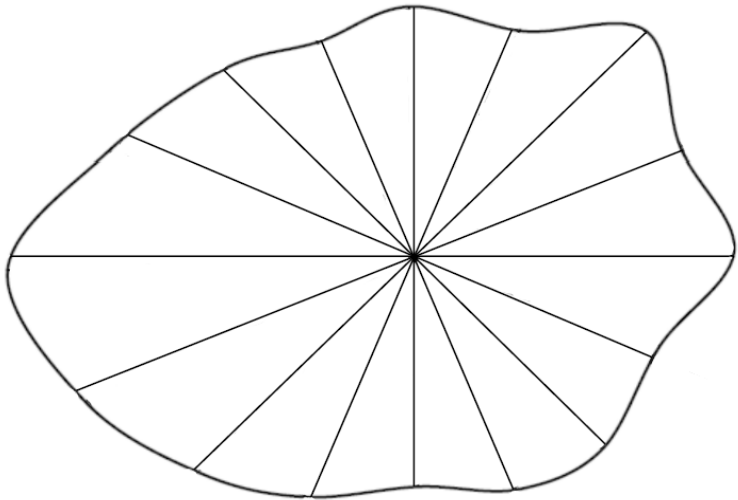

Figure 2     Brown's[3] protractor. Each interval shown is one sixteenth of a rotation. Further subdivisions can be added as desired.

Misconception:  the radian is a derived unit equal to a metre per metre

The SI Brochure specifies that the radian is to be treated as a derived unit equal to a metre per metre. This statement is refuted by the foregoing discussion, but we address it explicitly as well. Since it has been shown that angles can be measured in various ways, which would lead one to such alternative units for angles as $m^2/m^2$ or kg/kg, it is clear that the symbol m/m has no special connection to angles or the unit rad.  In fact, the apparent ambiguity in these units is due to the fact that all of these ratios are simply the number 1.  This follows from the SI prescription that the units follow the rules of algebra, one of which is that $x/x$ = 1.  The example of Brown's protractor shows that no other units need to be involved to define a measure of angles, and, as explained above, the idea that angles are inherently ratios is a misconception.

**Conclusion**

We believe that the reasoning given above shows definitively that angles are neither inherently length ratios, nor inherently dimensionless, and as a corollary that a radian is not a metre per metre.

It is then natural to ask: "but if angle is not a dimensionless ratio, what is it?" Returning to the bathroom scales analogy, we could equally well ask: "if mass is not a dimensionless ratio, what is it?" The convention is to treat mass as an independent quantity, and within the SI to give it a base unit and its own "dimension". This is not the only option, however. We could, as the current revision to the SI indirectly suggests, make action a base quantity, so that mass is a dimensional derived quantity with the units of (for example) action x time/length$^2$. Or we could treat mass as dimensionless, by choosing to set some particular mass equal to one by convention, as is done with the Hartree system of atomic units[4], the particular mass in that case being that of an electron. Treating mass as a dimensionless derived unit equal to a newton per newton is, however, not a reasonable option.

---

[3] The ruler-free protractor is named after its originator Richard Brown of NPL.
[4] In these units, $m_e$, $e$, $ℏ$ and $1/4πε_0$ are all set equal to unity.

The same options exist for angle. It could be treated as a base quantity, with its own base unit[5]. Alternatively, we could make angular momentum a base quantity, with the dimensions of energy x time/angle, and treat angle as a dimensional derived quantity. Or it can be treated as dimensionless, most conveniently by choosing to set $θ_N$ equal to one by convention. From the arguments in this paper, we conclude that there is no reasonable option for a radian to be a dimensionless derived unit equal to a metre per metre. None of the reasonable options corresponds to what is written in the SI Brochure.

We propose that the future treatment of angles within the SI should be based on consideration of reasonable options, not further promotion of the misconception that angles are inherently dimensionless or that the radian is equal to a metre per metre.

---

[5] If the concept of base unit becomes completely replaced by the "defining constant" approach, this option could be described by defining $θ_N$, the constant whose meaning is specified in the text above, as equal to exactly 1 radian.